# Performance Analysis of Linear Algebraic Functions using Reconfigurable Computing

Issam Damaj and Hassan Diab

issamwd@yahoo.com; diab@aub.edu.lb

**ABSTRACT:** This paper introduces a new mapping of geometrical transformation on the MorphoSys (M1) reconfigurable computing (RC) system. New mapping techniques for some linear algebraic functions are recalled. A new mapping for geometrical transformation operations is introduced and their performance on the M1 system is evaluated. The translation and scaling transformation addressed in this mapping employ some vector-vector and vector-scalar operations [6-7]. A performance analysis study of the M1 RC system is also presented to evaluate the efficiency of the algorithm execution. Numerical examples were simulated to validate our results, using the MorphoSys mULATE program, which emulates M1 operations.

## 1. INTRODUCTION

Reconfigurable computing (RC) is becoming more popular and increasing research efforts are being invested in it [1]. It employs reconfigurable hardware and programmable processors. The application is mapped such that the workload is divided between the general-purpose processor (GPP) and the reconfigurable device. The use of RC paves the way for an increased speed over general-purpose processors and a wider functionality than application specific integrated circuits (ASICs). It is a good solution for applications requiring a wide range of functionality and speed at the same time [1]. RC systems represent a solution to the inflexibility of ASICs on the one end of the computing spectrum, and the inefficiency of GPPs on the other end of the spectrum.

## 2. MORPHOSYS DESIGN

One of the emerging RC systems includes the MorphoSys designed and implemented at the University of California, Irvine. It has the block diagram shown in Figure 1 [2]. It is composed of: 1) an array of reconfigurable cells called the RC array, 2) its configuration data memory called context memory, 3) a control processor (TinyRISC), 4) a data buffer called the frame buffer, and 5) a DMA controller [2].

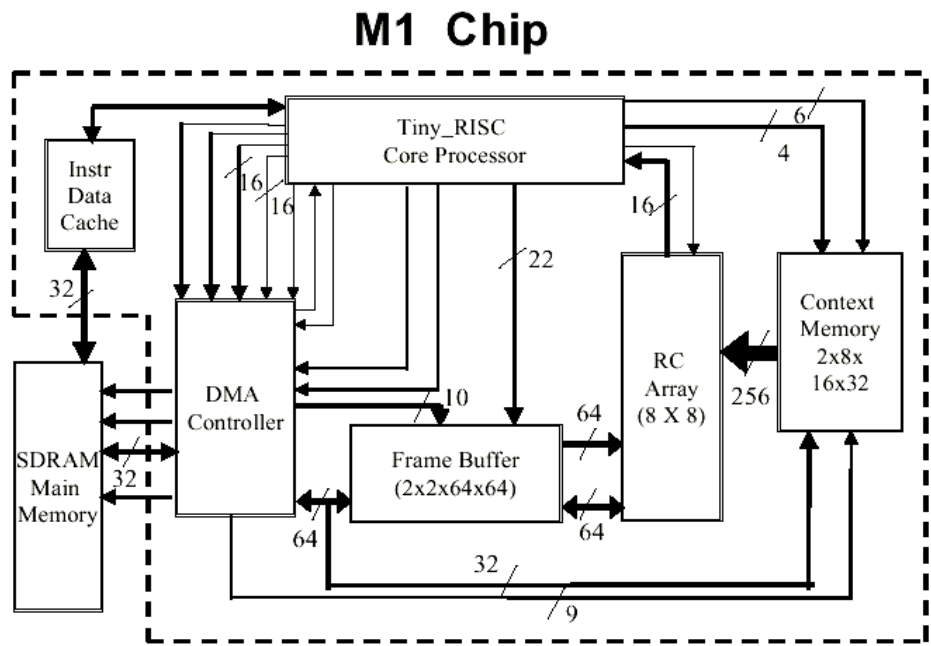

Figure 1.   MorphoSys block diagram

A program runs on MorphoSys in the following manner:  General-purpose operations are handled by the TinyRISC processor, while operations that have a certain degree of parallelism, regularity, or intensive computations are mapped to the RC array. The TinyRISC processor controls, through the DMA controller, the loading of the context words to context memory. These context words define the function and connectivity of the cells in the RC array. The processor also initiates the loading of the application data, such as image frames, from main memory to the frame buffer. This is also done through the DMA controller. Now that both configuration and application data are ready, the TinyRISC processor instructs the RC array to start execution. The RC array performs the needed operation on the application data and writes it back to the frame buffer. The RC array loads new application data from the frame buffer and possibly new configuration data from context memory. Since the frame buffer is divided into two sets, new application data can be loaded into it without interrupting the operation of the RC array. Configuration data is also loaded into context memory without interrupting RC array operation. This causes MorphoSys to achieve high speeds of execution [3].

### 3. RECONFIGURABLE DEVICE

As stated earlier, the reconfigurable device in MorphoSys is the RC array divided into four quadrants. It has the design and interconnection shown in Figure 2 [2]. The RC interconnection network is comprised of three hierarchical levels. The first layer provides a nearest neighbor connectivity that connects the RCs in a 2-D mesh. The second layer is an intra-quadrant connection that connects a specific RC to any other RC in its row or column in the same quadrant (a quadrant is a 4 by 4 group of cells in the RC array). The third layer is an inter-quadrant (or express lane) connection that carries data from any one cell (out of four) in a row (or column) of a quadrant to other cells in an adjacent quadrant but in the same row (or column) [4].

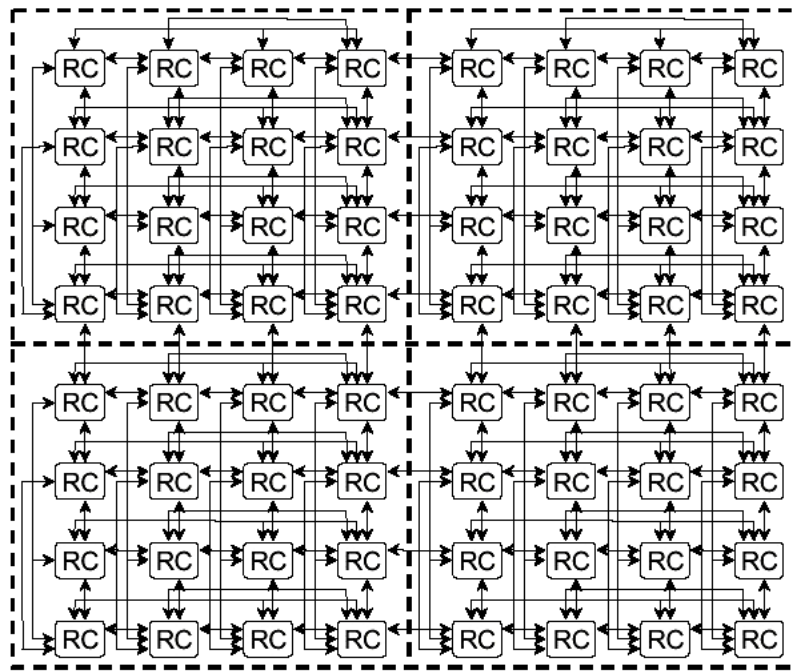

Figure 2.  RC array interconnection.

The context words present on context memory configure the function of the RCs as well as the interconnection, thus specifying where their input is from and where their output will be written [5]. MorphoSys is designed in a way where all the cells in the same row perform the same function and have the same connection scheme (in row context broadcast mode), or all the cells in the same column perform the same function and have the same connections scheme (in column context broadcast mode). All the cells of a row or of a column share the same configuration word [5].

The *reconfigurable cell* is the basic programmable element in MorphoSys. Each *reconfigurable cell* (Figure 3) comprises five components: the ALU/Multiplier, the shift unit, the input multiplexers, a register file with four-bit registers, and the context register. There are 64 *reconfigurable cells* arranged as an 8x8 matrix called the RC Array. The ALU/Multiplier has four data input ports: two 16-bit ports receive data from the input multiplexers, one bit port takes data from the output register, and a bit port takes an immediate value in the context word. In addition to standard arithmetic and logical operations, the ALU/Multiplier can perform a multiply accumulate operation in a single cycle. The shift unit is also 32-bits wide.

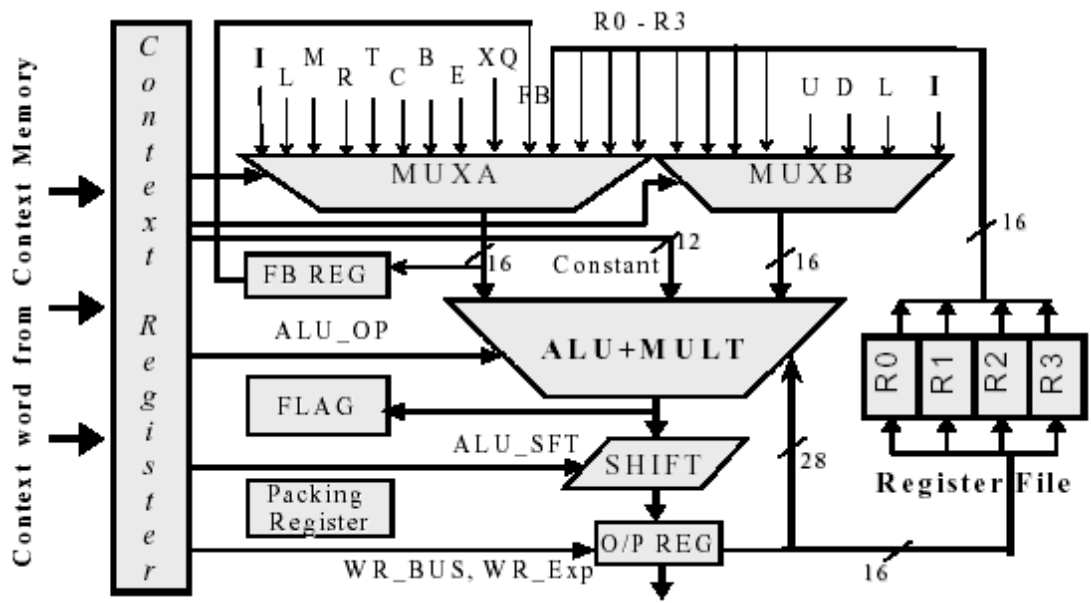

Figure 3.   Architecture of a reconfigurable cell.

In the current MorphoSys prototype, the ALU-Multiplier operates only on signed numbers. However, several important applications such as data encryption/decryption involve multiplication of unsigned numbers. Therefore, the ALU-Multiplier will be extended for operation using both signed and unsigned values in the next implementation of MorphoSys.

The input multiplexers (Figure 3) select one of several inputs for the ALU/Multiplier Multiplexer A selects one input from: (a) four nearest neighbors in the RC Array, or (b) other RCs in the same row/column within the same RC Array quadrant, or (c) the operand data bus, or (d) the internal register file. Multiplexer B selects one input from: (a) three of the nearest neighbors, or (b) the operand bus, or (c) the register file.

The context register provides control signals for the RC components through the context word. The bits of the context word directly control the input multiplexers the ALU/Multiplier and the shift unit. The context word determines the destination of a result, which can be a register in the register file and/or the express lane (64-bit) buses. The context word also has a field for an immediate operand value.

## 4. GEOMETRICAL TRANSFORMATIONS IN COMPUTER GRAPHICS

Transformations are a fundamental part of computer graphics. Transformations are used to position, shape, and change viewing positions of objects, as well as change how they are viewed (e.g. the type of projection used). There are many types of two-dimensional transformations that one can perform; however, in this paper we will address: Translation and Scaling. These basic transformations can also be combined to obtain more complex transformations. Figure 4 shows the effects of some 2D transformations on an image.

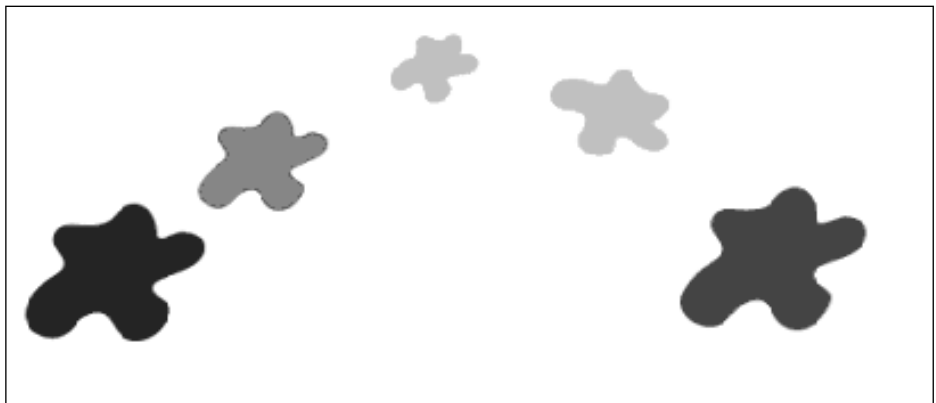

Figure 4.  Image tracking while applying different 2D transformations.

A point p in 2D is represented by p(x, y) where x is the x-coordinate and y is the y-coordinate of p. 2D objects are often represented as a set of points (vertices), $\{P_1, P_2, ..., P_n\}$, and an associated set of edges $\{e_1, e_2, ..., e_m\}$. An edge is defined as a pair of points, $e\{P_i, P_j\}$. We can also represent points in vector/matrix notation as:

$$\mathbf{P} = \begin{bmatrix} \mathbf{x} \\ \mathbf{y} \end{bmatrix}$$

*Translations*

A translation can also be represented by a pair of numbers, $t=(t_x, t_y)$ where $t_x$ is the change in the x-coordinate and $t_y$ is the change in y-coordinate. To translate the point p by t, we simply add to obtain the new (translated) point $q(x', y') = p(x, y) + t(t_x, t_y)$.

$$\mathbf{q} = \begin{bmatrix} \mathbf{x} \\ \mathbf{y} \end{bmatrix} + \begin{bmatrix} \mathbf{t_x} \\ \mathbf{t_y} \end{bmatrix} = \begin{bmatrix} \mathbf{x} + \mathbf{t_x} \\ \mathbf{y} + \mathbf{t_y} \end{bmatrix}$$

Figure 5 is an example of translation on a simple image.

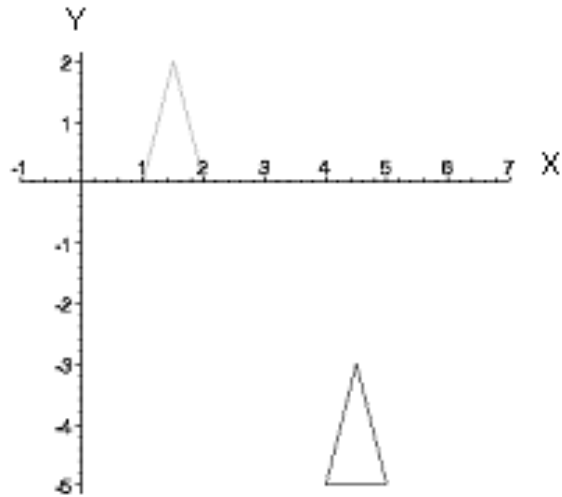

Figure 5. Translation

## Scaling

Matrices can easily represent scaling transformations. Let the scale matrix be "s", and the new point "q".

$$s = \begin{bmatrix} s_x & 0 \\ 0 & s_y \end{bmatrix}; \quad q = s \times p = \begin{bmatrix} s_x & 0 \\ 0 & s_y \end{bmatrix} \times \begin{bmatrix} x \\ y \end{bmatrix}$$

Where $q = \begin{bmatrix} x' \\ y' \end{bmatrix} = \begin{bmatrix} s_x x \\ s_y y \end{bmatrix}$

Figure 6 is an example of scaling on a simple image. Both translation and scaling are transformations applied with respect to the origin. Therefore, Figure 6 shows an inherent translation when applying a scaling transformation.

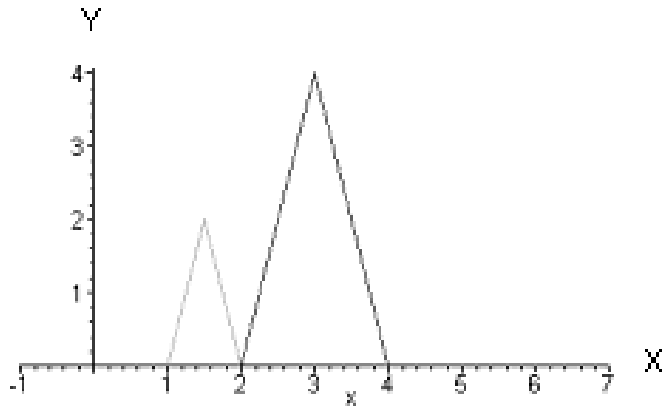

Figure 6. Scaling

## 5. ALGORITHM MAPPING

The main usage of the MorphoSys is, as any parallel processing system, to perform fast computations of algorithms that need a certain computational power requirement. Computer graphics algorithms represent one of these families. Computer graphics accelerators are the subject of much research. A basic part in computer graphics operations is geometrical transformations, which require fast computations of vector-vector operations and vector-scalar operations. The emphasis in this paper is the mapping of vector and scalar operations on the MorphoSys, e.g. vectors addition, subtraction, and multiplication by a scalar. These algorithms represent the core of many algorithms, especially, those used for translation and scaling in geometrical transformations.

### 5.1 TRANSFORMATION WITH VECTORS OPERATIONS

Generally, a one-dimensional n-element vector has the form: $M^T = [M_0 M_1 M_2 ..... M_{n-1}]$. In our case of geometrical transformation, a vector U could be considered as the original coordinates, while a vector V could be considered as the corresponding translation values. Mapping an algorithm for addition, or any other operation, of the two vectors is done by first storing them in the Frame Buffer set "0" and set "1". Then we can exploit the properties of the interconnection, where some contents of Frame Buffer set "0" are added to some contents of Frame Buffer set "1" and the result would be in columns 0-7 of the RC-array. Figure 7 shows the final output in the RC-array after running the algorithm of adding two 64-element vectors.

| Columns Rows | $C_0$ | $C_1$ | $C_2$ | $C_3$ | $C_4$ | $C_5$ | $C_6$ | $C_7$ |
|---|---|---|---|---|---|---|---|---|
| $R_0$ | $U_0+V_0$ | $U_8+V_8$ | $U_{16}+V_{16}$ | $U_{24}+V_{24}$ | $U_{32}+V_{32}$ | $U_{40}+V_{40}$ | $U_{48}+V_{48}$ | $U_{56}+V_{56}$ |
| $R_1$ | $U_1+V_1$ | $U_9+V_9$ | $U_{17}+V_{17}$ | $U_{25}+V_{25}$ | $U_{33}+V_{33}$ | $U_{41}+V_{41}$ | $U_{49}+V_{49}$ | $U_{57}+V_{57}$ |
| $R_2$ | $U_2+V_2$ | $U_{10}+V_{10}$ | $U_{18}+V_{18}$ | $U_{26}+V_{26}$ | $U_{34}+V_{34}$ | $U_{42}+V_{42}$ | $U_{50}+V_{50}$ | $U_{58}+V_{58}$ |
| $R_3$ | $U_3+V_3$ | $U_{11}+V_{11}$ | $U_{19}+V_{19}$ | $U_{27}+V_{27}$ | $U_{35}+V_{35}$ | $U_{43}+V_{43}$ | $U_{51}+V_{51}$ | $U_{59}+V_{59}$ |
| $R_4$ | $U_4+V_4$ | $U_{12}+V_{12}$ | $U_{20}+V_{20}$ | $U_{28}+V_{28}$ | $U_{36}+V_{36}$ | $U_{44}+V_{44}$ | $U_{52}+V_{52}$ | $U_{60}+V_{60}$ |
| $R_5$ | $U_5+V_5$ | $U_{13}+V_{13}$ | $U_{21}+V_{21}$ | $U_{29}+V_{29}$ | $U_{37}+V_{37}$ | $U_{45}+V_{45}$ | $U_{53}+V_{53}$ | $U_{61}+V_{61}$ |
| $R_6$ | $U_6+V_6$ | $U_{14}+V_{14}$ | $U_{22}+V_{22}$ | $U_{30}+V_{30}$ | $U_{38}+V_{38}$ | $U_{46}+V_{46}$ | $U_{54}+V_{54}$ | $U_{62}+V_{62}$ |
| $R_7$ | $U_7+V_7$ | $U_{15}+V_{15}$ | $U_{23}+V_{23}$ | $U_{31}+V_{31}$ | $U_{39}+V_{39}$ | $U_{47}+V_{47}$ | $U_{55}+V_{55}$ | $U_{63}+V_{63}$ |

Figure 7. RC array contents after matrix addition

For the MorphoSys to perform the required calculations, three sets of data must be first entered to the M1 chip. Firstly, the TinyRISC program that controls the functionality of the whole system. This code is placed in main memory (Figure 1) and handles all the operations that are not mapped onto the RC array such as data transfer. It also provides the RC array with which contexts it needs to run, what data to access as input, and where the output data is written. Secondly, the code for the RC array operation is called the context code. This is written for either column mode or row mode or for both. It defines what operation each row or column is going to carry out, what input it takes, and where the output is to be stored. Finally, the data required for computations is stored in the Frame Buffer for later retrieval and use by the TinyRISC program.

Let the desired function of the interconnection be: *Out = A + B*. Thus, the context word would be: 0000F400. This must be loaded into the column block of the context memory. Assume that vector U is stored in address 10,000$_{hex}$ of main memory, vector V stored in address 20,000$_{hex}$ and the context word stored in address 30,000$_{hex}$. Then the answer will be stored back to address 40,000$_{hex}$. The M1 code and its discussion are provided in Table 1.

Table 1.   TinyRISC code for the uniform translation routine for a 64-element vector.

| 0: | ldui | r1, 0x1; | R1 ← 10000$_{hex}$. Vector U is stored. |
|---|---|---|---|
| 1: | ldfb | r1, 0, 0, 16 ; | FB ← 16 x 32 bits at set 0, bank A, address 0. |
| 2: | add | r0, r0, r0; | No-operation. |
| . | . | . | |
| 33: | ldui | r1, 0x2; | R1 ← 20000$_{hex}$.<br>This is where vector V is stored. |
| 34: | ldfb | r1, 1, 0, 16; | FB ← 16 x 32 bits at set 0, bank B, address 0. |
| 35: | add | r0, r0, r0; | NOP |
| . | . | | |
| 66: | ldui | r3, 0x3; | R3 ← 30000$_{hex}$. This is where the context word is stored in main memory. |
| 67: | ldctxt | r3, 0, 0, 0, 1; | Load one context word from main memory starting at the address stored in register 3 into plane 0, block 0 and starting at word 0. |
| 68: | add | r0, r0, r0; | NOP |
| . | . | . | |
| 71: | ldui | r4, 0x0; | R4 ← 00000$_{hex}$. |
| 72: | dbcdc | r4, 0, 0, 0, 0, 0, 0; | Double bank column broadcast. It sends data from both banks address 0 in the frame buffer and broadcasts the context words column-wise. It triggers the RC array to start execution of column 0 by the context word of address 0 in the column block of context memory operating on data in set 0. Bank A starting at 0x0. Bank B starting at (0x0 + 0). |
| 73: | ldli | r4, 0x4 | R4 ← 4$_{hex}$ |
| 74: | dbcdc | r4, 0, 0, 1, 0, 0, 0x40; | It sends data from both banks address 40$_{hex}$ in the frame buffer. Bank A starting at 0x40. Bank B starting at (0x4 + 0x0 = 0x40). |
| 75: | ldli | r4, 0x8 | R4 ← 8$_{hex}$ |
| 76: | dbcdc | r4, 0, 0, 2, 0, 0, 0x80; | It sends data from both banks. |
| 77: | ldli | r4, 0xC | R4 ← C$_{hex}$ |
| 78: | dbcdc | r4, 0, 0, 3, 0, 0, 0xC0; | It sends data from both banks address C0$_{hex}$ in the frame buffer. Bank A starting at 0xC0. Bank B starting at (0xC + 0x0 = 0xC0). |
| 79: | ldli | r4, 0x10 | R4 ← 10$_{hex}$ |
| 80: | dbcdc | r4, 0, 0, 4, 0, 0, 0x100; | It sends data from both banks address 100$_{hex}$ in the frame buffer. Bank A starting at 0x100. Bank B starting at (0x10 + 0x0 = 0x100). |
| 81: | ldli | r4, 0x14 | R4 ← 14$_{hex}$ |
| 82: | dbcdc | r4, 0, 0, 5, 0, 0, 0x140; | It sends data from both banks address 140$_{hex}$ in the frame buffer. Bank A starting at 0x140. Bank B starting at (0x14 + 0x0 = 0x140). |
| 83: | ldli | r4, 0x18 | R4 ← 18$_{hex}$ |
| 84: | dbcdc | r4, 0, 0, 6, 0, 0, 0x180; | It sends data from both banks. |
| 85: | ldli | r4, 0x1C | R4 ← 1C$_{hex}$ |
| 86: | dbcdc | r4, 0, 0, 7, 0, 0, 0x1C0; | It sends data from both banks. |
| 87: | wfbi | 0, 0, 0, 1, 0x0; | Write data back to the frame buffer from the output registers<br>    of column 0 into set 1, address 0. |
| 88: | wfbi | 1, 0, 0, 1, 0x40; | of column 1 into set 1, address 64. |
| 89: | wfbi | 2, 0, 0, 1, 0x80; | of column 2 into set 1, address 128. |
| 90: | wfbi | 3, 0, 0, 1, 0xC0; | of column 3 into set 1, address 192. |
| 91: | wfbi | 4, 0, 0, 1, 0x100; | of column 4 into set 1, address 256. |
| 92: | wfbi | 5, 0, 0, 1, 0x140; | of column 5 into set 1, address 320. |
| 93: | wfbi | 6, 0, 0, 1, 0x180; | of column 6 into set 1, address 384. |
| 94: | wfbi | 7, 0, 0, 1, 0x1C0; | of column 7 into set 1, address 448. |
| 95: | ldui | r5, 0x4; | R5 ← 40000hex. |
| 96: | stfb | r1, 1, 0,10$_{hex}$; | Store data from frame buffer set 1, address 0 into main memory starting at address stored in reg1. |

## 5.2. TRANSFORMATION WITH VECTORS & SCALARS OPERATIONS

Consider an 8-element vector U.

$$U^T = [U_0 \ U_1 \ U_2 \ U_3 \ U_4 \ U_5 \ U_6 \ U_7]$$
$$W = c \times U \Rightarrow \ W^T = [cU_0 \ cU_1 \ .... cU_7]$$

For n-element vectors: $W^T = [cU_0 \ cU_1 ..... cU_n]$.

Mapping the algorithm for multiplication, or any other operation (arithmetic or logical), of a vector by a scalar, is done by first storing the vector in the Frame Buffer set "0". Then we can exploit the properties of the interconnection, where some contents of the Frame Buffer set "0" are multiplied by a constant to be stored in the context word. Figure 8 shows the final output in the RC-array after running the algorithms of two 64-element vectors.

| Columns Rows | $C_0$ | $C_1$ | $C_2$ | $C_3$ | $C_4$ | $C_5$ | $C_6$ | $C_7$ |
|---|---|---|---|---|---|---|---|---|
| $R_0$ | $c.U_0$ | $c.U_8$ | $c.U_{16}$ | $c.U_{24}$ | $c.U_{32}$ | $c.U_{40}$ | $c.U_{48}$ | $c.U_{56}$ |
| $R_1$ | $c.U_1$ | $c.U_9$ | $c.U_{17}$ | $c.U_{25}$ | $c.U_{33}$ | $c.U_{41}$ | $c.U_{49}$ | $c.U_{57}$ |
| $R_2$ | $c.U_2$ | $c.U_{10}$ | $c.U_{18}$ | $c.U_{26}$ | $c.U_{34}$ | $c.U_{42}$ | $c.U_{50}$ | $c.U_{58}$ |
| $R_3$ | $c.U_3$ | $c.U_{11}$ | $c.U_{19}$ | $c.U_{27}$ | $c.U_{35}$ | $c.U_{43}$ | $c.U_{51}$ | $c.U_{59}$ |
| $R_4$ | $c.U_4$ | $c.U_{12}$ | $c.U_{20}$ | $c.U_{28}$ | $c.U_{36}$ | $c.U_{44}$ | $c.U_{52}$ | $c.U_{60}$ |
| $R_5$ | $c.U_5$ | $c.U_{13}$ | $c.U_{21}$ | $c.U_{29}$ | $c.U_{37}$ | $c.U_{45}$ | $c.U_{53}$ | $c.U_{61}$ |
| $R_6$ | $c.U_6$ | $c.U_{14}$ | $c.U_{22}$ | $c.U_{30}$ | $c.U_{38}$ | $c.U_{46}$ | $c.U_{54}$ | $c.U_{62}$ |
| $R_7$ | $c.U_7$ | $c.U_{15}$ | $c.U_{23}$ | $c.U_{31}$ | $c.U_{39}$ | $c.U_{47}$ | $c.U_{55}$ | $c.U_{63}$ |

Figure 8.   RC array contents after multiplication by a scalar of a 64-element vector is performed

The desired function of the interconnection is: *Out (t+1) = c x A*. Lets assume that the constant "c" has a value of $5_{hex}$. According to the internal bit assignment of the ALU control lines, the context word is:  00009005. This word must be loaded into the column block of the context memory. Assume that vector U is stored in address $30,000_{hex}$ of main memory, and the context word stored in address $40,000_{hex}$. Then the answer will be stored back to address $50,000_{hex}$. The M1 code and its discussion are provided in Table 2.

Table 2.   TinyRISC code for the uniform scaling routine of a 64-element vector or an 8x8 matrix.

| 0: | ldui | r1, 0x1; | R1 ← 10000$_{hex}$. This is where vector U is stored. |
|---|---|---|---|
| 1: | ldfb | r1, 0, 0, 16 ; | FB ← 16 x 32 bits at set 0, bank A, address 0. |
| 2: | Add | r0, r0, r0; | No-operation. |
| . | . | . | |
| 33: | ldui | r3, 0x3; | R3 ← 30000$_{hex}$. This is where the context word is stored in main memory. |
| 34: | ldctxt | r3, 0, 0, 0, 1; | Load one context word from main memory starting at the address stored in register 3 into plane 0, block 0 and starting at word 0. |
| 35: | add | r0, r0, r0; | NOP |
| . | . | . | |
| 38: | sbcb | 1, 0, 0, 0, 0, 0, 0x0; | Single bank column broadcast causing all the cells in the RC array to perform their operations specified by the context word in context memory starting with data from frame buffer, set 0, bank A, address offset 0. |
| 39: | sbcb | 1, 0, 0, 0, 0, 0, 0x40; | It sends data from both banks address 40$_{hex}$ in FB. |
| 40: | sbcb | 1, 0, 0, 0, 0, 0, 0x80; | It sends data from both banks address 80$_{hex}$ in FB. |
| 41: | sbcb | 1, 0, 0, 0, 0, 0, 0xC0; | It sends data from both banks address C0$_{hex}$ in FB |
| 42: | sbcb | 1, 0, 0, 0, 0, 0, 0x100; | It sends data from both banks address 100$_{hex}$ in the frame buffer. |
| 43: | sbcb | 1, 0, 0, 0, 0, 0, 0x140; | It sends data from both banks address 140$_{hex}$ in the frame buffer. |
| 44: | sbcb | 1, 0, 0, 0, 0, 0, 0x180; | It sends data from both banks address 180$_{hex}$ in the frame buffer. |
| 45: | sbcb | 1, 0, 0, 0, 0, 0, 0x1C0; | It sends data from both banks address 1C0$_{hex}$ in the frame buffer. |
| 46: | wfbi | 0, 0, 0, 1, 0x0; | Write data back to the frame buffer from the output registers of column 0 into set 1, address 0. |
| 47: | wfbi | 1, 0, 0, 1, 0x40; | of column 1 into set 1, address 64. |
| 48: | wfbi | 2, 0, 0, 1, 0x80; | of column 2 into set 1, address 128. |
| 49: | wfbi | 3, 0, 0, 1, 0xC0; | of column 3 into set 1, address 192. |
| 50: | wfbi | 4, 0, 0, 1, 0x100; | of column 4 into set 1, address 256. |
| 51: | wfbi | 5, 0, 0, 1, 0x140; | of column 5 into set 1, address 320. |
| 52: | wfbi | 6, 0, 0, 1, 0x180; | of column 6 into set 1, address 384. |
| 53: | wfbi | 7, 0, 0, 1, 0x1C0; | of column 7 into set 1, address 448. |
| 54: | ldui | r5, 0x4; | R5 ← 40000hex. |
| 55: | stfb | r1, 1, 0, 10$_{hex}$; | Store data from frame buffer set 1, address 0 into main memory starting at address stored in reg1. |

## 5.3. ROTATION TRANSFORMATIONS

Rotation and composite transformations represent two other kinds of graphics geometrical transformations. Both transformations are mapped onto the MorphoSys as a matrix multiplication operation [8]. Results are presented in the following section. Multiplying matrix A with matrix B would mean multiplying row one ($r_1$) of A with column one of B and then adding their results yielding ($c_{11}$) of the result matrix C. The multiplication with ($r_1$) is repeated to all columns of B yielding ($c_2$ .. $c_n$). Then, ($r_2$) of A repeats the same multiplication with all columns of B. This algorithm is repeated until the last row in A.

Matrices A, B, and C are dense matrices. The matrix-matrix multiplication involves $O(n^3)$ operations on a single processing plat form, since for each element $C_{ij}$ of C, we must compute $C_{ij} = \sum_{k=0}^{N-1} A_{ik} B_{kj}$. This simple Algorithm is mapped onto the M1 RC-Array as follows: The contents of matrix A are passed row by row through the context words, thus, stored in the context memory for later retrieval and manipulation by the reconfigurable cells. The contents of matrix B are broadcasted also row by row to the columns of the RC array. The multiplication stage (row x column) is done by using the CMUL (constant-multiply) ALU operation where Out (t) = AxB; which is the required computation. Note that CMUL is a vector-scalar operation discussed in [7].

## 6. PERFORMANCE ANALYSIS

Performance evaluation and comparisons are made among the mappings already suggested. This performance is based on the speed of execution of the algorithms. The MorphoSys M1 system is operational at a frequency of 100 MHz. After obtaining the results of the mapped algorithms onto the M1 system, a comparison is done with the mapping of the algorithms onto single-processor systems namely the Intel 80386 and 80486. These processors where chosen as they are comparable in CPU speed to the M1 system. Note that the instructions used are upward compatible with later versions of the Intel processors.

### 6.1. PERFORMANCE OF THE TRANSLATION ALGORITHMS

The code using the Intel Instruction set is shown in Table 3. The comparisons between the three different processing systems are shown in Figures 9-12. These figures clearly show the superiority of MorphoSys over the other suggested processors. The number of elements per cycle for the 8-element vector-vector translation algorithms yielded the following ratios (in elements/cycle) on the M1, 80386, and the 80486: 0.38, 0.036, and 0.088 respectively. For the case of 64-element vectors it yielded: 0.667, 0.036, and 0.083 respectively. Therefore, the performance of M1 compared to the 80486/386 is superior with respect to the number elements that can be processed per cycle.

Table 3.    Code and analysis of vector-vector algorithms, used for translation, using the Intel 80386, and 80486 instructions set.

| Label | | | Description | Clocks | |
|---|---|---|---|---|---|
| | | | | 80486 | 80386 |
| | MOV | SP, V1_Loc | SP ← Location of V1 in memory. | 1T | 2T |
| | MOV | BP, V2_Loc | BP ← Loc. of V2 in memory. | 1T | 2T |
| | MOV | DI, Result_Loc | SP ← Loc. of the result in mem. | 1T | 2T |
| | MOV | SI, Count_Value | SI ← Value to count down. 8 for Algorithm in table. And 64 for the algorithm in table. | 1T | 2T |
| AA: | MOV | AX, [SP] | Get value addressed by the SP | 1T | 4T |
| | MOV | BX, [BP] | Get value addressed by BP reg. | 1T | 4T |
| | ADD | AX, BX | AX ← AX + BX. | 1T | 2T |
| | MOV | [DI], AX | Store the value in AX in the memory location addressed by DI register. | 1T | 2T |
| | INC | SP | Get next element of V1. | 1T | 2T |
| | INC | BP | Get next element of V2. | 1T | 2T |
| | INC | DI | Increment the destination. | 1T | 2T |
| | DEC | SI | Decrement the counter. | 1T | 2T |
| | JNZ | AA | Jump if not finished to label AA. | 3/1T | 7/3T |
| Calculations | | | **TOTAL Time States** | | |
| | | | For an 8-element vector | 90T | 220T |
| | | | For a 64-element vector | 769T | 1723T |
| | | | Frequency | 100MHz | 40MHz |
| | | | For an 8-element vector | 0.9μs | 5.5μs |
| | | | For a 64-element vector | 7.69μs | 43.075μs |
| | | | | | |
| | | | **Elements Per Cycle** | | |
| | | | For an 8-element vector | 0.088 | 0.036 |
| | | | For a 64-element vector | 0.083 | 0.037 |
| | | | | | |
| | | | **Cycles Per Elements** | | |
| | | | For an 8-element vector | 11.36 | 27.5 |
| | | | For a 64-element vector | 12 | 26.9 |

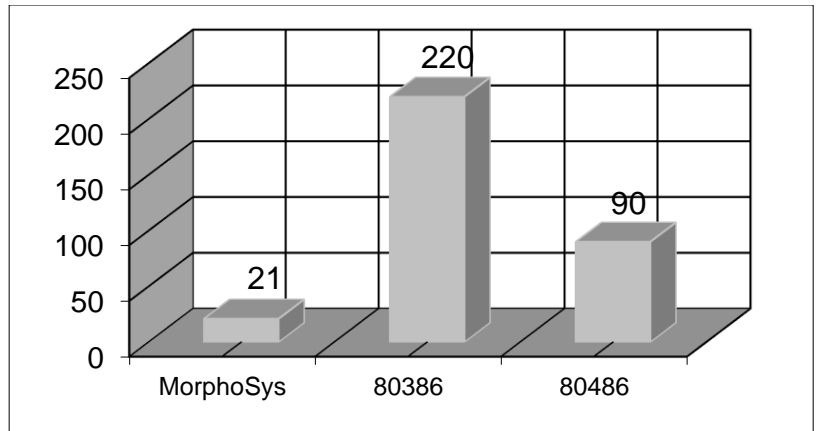

Figure 9.  Comparison wrt number of cycles taken to run the 8-element vector-vector translation algorithm.

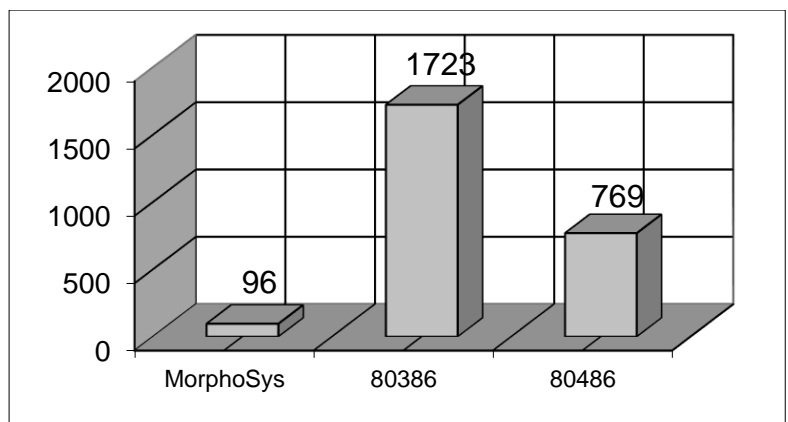

Figure 10.  Comparison wrt number of cycles taken to run the 64-element vector-vector translation algorithm.

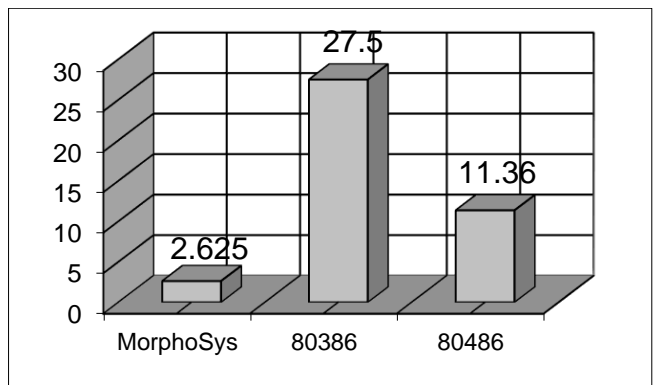

Figure 11.  Comparison wrt number of cycles/element to run the 8-element vector-vector translation algorithm.

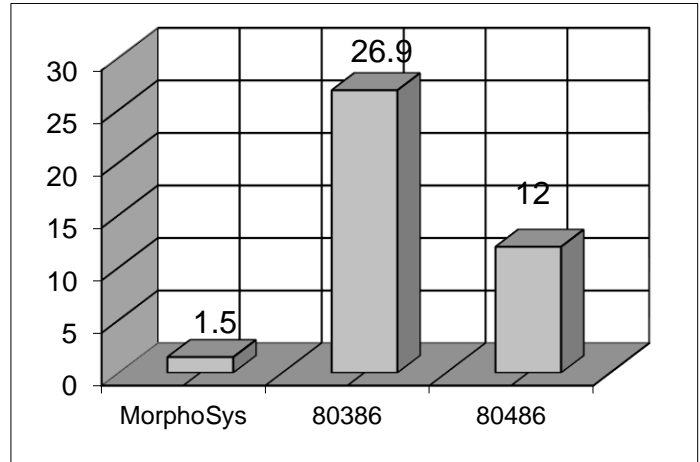

Figure 12. Comparison wrt number of cycles/element to run the 64-element vector-vector translation algorithm.

## 6.2 PERFORMANCE OF THE SCALING ALGORITHMS

The code using the Intel Instruction set is shown in Table 4. The comparisons between the three different processing systems are shown in Figures 13-16. These figures also clearly show the superiority of the MorphoSys over the other processor. The number of elements per cycle for the 8-element vector-vector translation algorithms yielded the following ratios (in elements/cycles) on the M1, 80386, and the 80486: 0.57, 0.046, and 0.108 respectively. For the case of 64-element vectors it yielded: 1.16, 0.046, and 0.11 respectively. Accordingly, the performance of M1 compared to the 80486/386 is superior with respect to the number elements that can be processed per cycle.

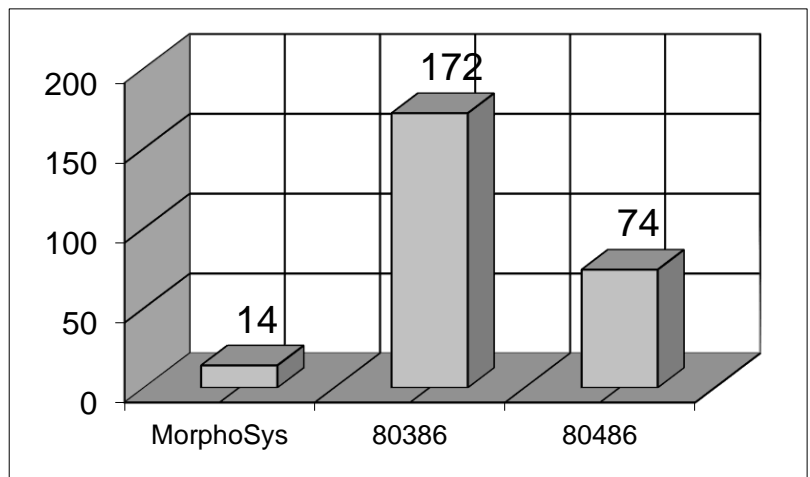

Figure 13. Comparison wrt number of cycles to run the 8-element vector-scalar scaling algorithm.

Table 4.   Code and analysis of vector-scalar algorithms, used for scaling, using the Intel 80386 and 80486 instructions set.

| Label | | | Description | Clocks | |
|---|---|---|---|---|---|
| | | | | 80486 | 80386 |
| | MOV | SP, V1_Loc | SP ← Location of V1 in mem. | 1T | 2T |
| | MOV | BP, Constant | BP ← Constant Scalar | 1T | 2T |
| | MOV | DI, Result_Loc | SP ← Loc of the result in mem. | 1T | 2T |
| | MOV | SI, Count_Value | SI ← Value to count down. 8 for Algorithm in table1. And 64 for that in table2. | 1T | 2T |
| AA: | MOV | AX, [SP] | Get the vector element addressed by the SP register. | 1T | 4T |
| | ADD | AX, BP | AX ← AX + Constant. | 1T | 2T |
| | MOV | [DI], AX | Store the value in AX in the memory location addressed by DI reg. | 1T | 2T |
| | INC | SP | Get next element of V1. | 1T | 2T |
| | INC | DI | Increment the destination. | 1T | 2T |
| | DEC | SI | Decrement the counter. | 1T | 2T |
| | JNZ | AA | Jump of not finished to label AA. | 3/1T | 7/3T |
| Calculations | | **TOTAL Time States** | | | |
| | | For an 8-element vector | | 74T | 172T |
| | | For a 64-element vector | | 578T | 1348T |
| | | Frequency | | 100MHz | 40MHz |
| | | For an 8-element vector | | 0.74µs | 4.3µs |
| | | For a 64-element vector | | 5.78µs | 33.7µs |
| | | **Elements Per Cycle** | | | |
| | | For an 8-element vector | | 0.108 | 0.046 |
| | | For a 64-element vector | | 0.11 | 0.047 |
| | | **Cycles Per Elements** | | | |
| | | For an 8-element vector | | 9.25 | 21.7 |
| | | For a 64-element vector | | 9.03 | 21.2 |

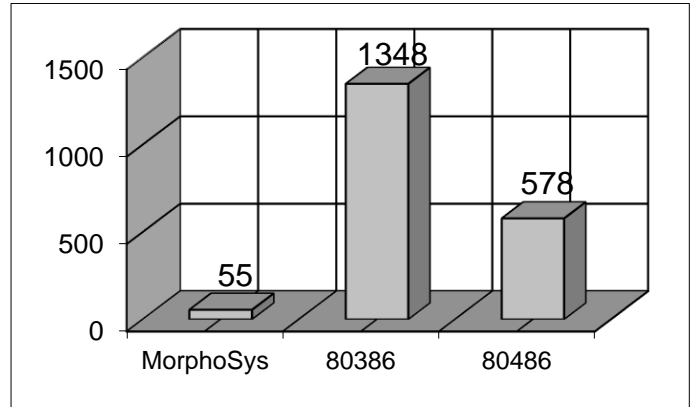

Figure 14. Comparison wrt number of cycles taken to run the 64-element vector-scalar scaling algorithm.

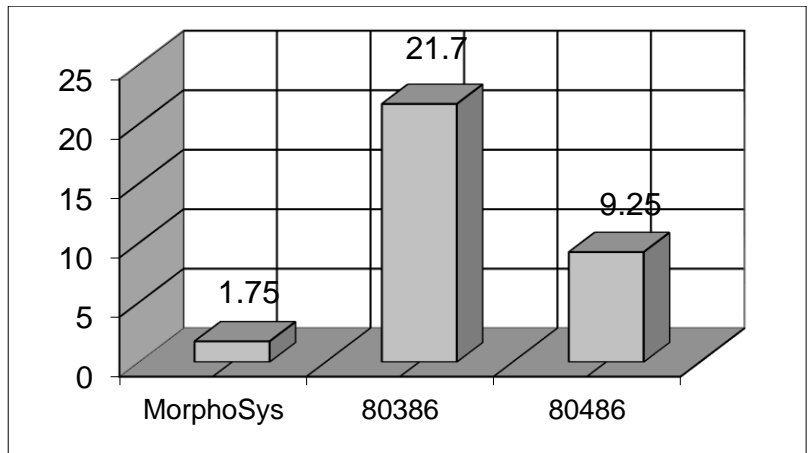

Figure 15. Comparison wrt number of cycles/element to run the 8-element vector-scalar scaling algorithm

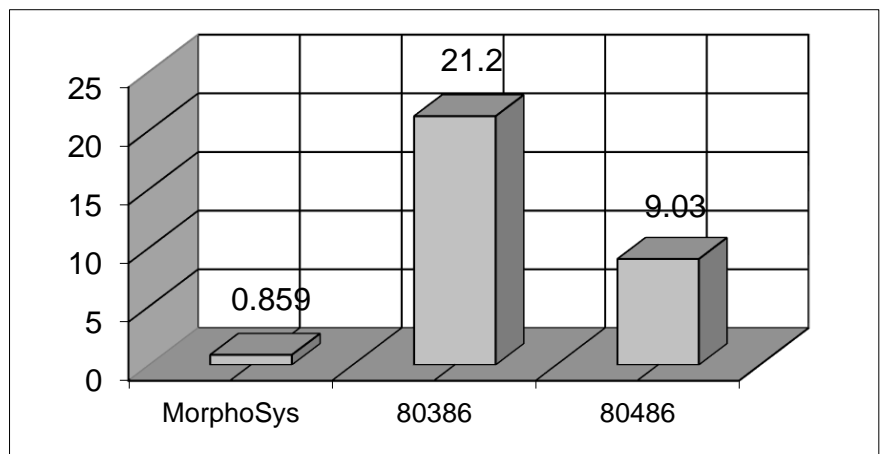

Figure 16. Comparison wrt number of cycles/element to run the 64-element vector-scalar scaling algorithm.

## 7. ANALYSIS OF RESULTS

Results in Table 5 show the superiority of the addressed RC-system the M1 over the other suggested systems. This is clearly indicated by the speedup factor that reached 24 in some cases. On the other hand, the major issue is not only comparing the suggested algorithms with other systems running the same algorithm, but it is the trials to find and tune the best possible algorithm mapping that gives the most favourable performance using this state-of-the-art reconfigurable M1 system. The speedup factors, shown in Table 5, are the ratios of the number of execution cycles of the M1 over the other systems. The discussed findings are part of a complete graphics acceleration library using the M1 reconfigurable system.

Table 5. Comparisons between different algorithms and systems. Clock speeds for the 80386, 80486, and Pentium are: 40, 100, and 133MHz respectively.

| Algorithm | System | No. Elements | No. of Cycles | Speedup | Total Time in Micro-Secs. | Elements per Cycle | Cycles per Element |
|---|---|---|---|---|---|---|---|
| Vector-Vector Operations (**Translation**) | M1 | 64 | 96 | | 0.96 | 0.667 | 1.5 |
| | 80486 | | 769 | **8.01** | 7.69 | 0.083 | 12 |
| | 80386 | | 1723 | **17.94** | 43.075 | 0.037 | 26.9 |
| Vector-Scaling Operations (**Scaling**) | M1 | 64 | 55 | | 0.55 | 1.16 | 0.859 |
| | 80486 | | 578 | **10.51** | 5.78 | 0.047 | 9.03 |
| | 80386 | | 1348 | **24.51** | 33.7 | 0.11 | 21.2 |
| General Composite Algorithm I Using Matrix Algorithm (**Rotation**) | M1 | 64 | 256 | | 2.56 | 0.25 | 4 |
| | Pentium | | 10151 | **39.65** | 76.32 | 0.006 | 158.6 |
| | 80486 | | 27038 | **105.62** | 270.38 | 0.002 | 422.4 |
| General Composite Algorithm II Using Matrix Algorithm (**Rotation**) | M1 | 16 | 70 | | 0.7 | 0.228 | 4.375 |
| | Pentium | | 1328 | **18.97** | 9.98 | 0.012 | 83 |
| | 80486 | | 3354 | **47.91** | 33.54 | 0.0047 | 209.6 |
| Vector-Vector Operations (**Translation**) | M1 | 8 | 21 | | 0.21 | 0.38 | 2.625 |
| | 80486 | | 90 | **4.29** | 0.9 | 0.088 | 11.36 |
| | 80386 | | 220 | **10.48** | 5.5 | 0.036 | 27.5 |
| Vector-Scaling Operations (**Scaling**) | M1 | 8 | 14 | | 0.14 | 0.57 | 1.75 |
| | 80486 | | 74 | **5.28** | 0.74 | 0.108 | 9.25 |
| | 80386 | | 172 | **12.29** | 4.3 | 0.46 | 21.7 |

## 8. CONCLUSION

The MorphoSys RC system has been utilized for different areas of application. With computer graphics computationally intensive algorithms, the M1 system has been used in image processing, graphics acceleration for animation, and currently with geometrical transformations. In this paper, new mapping techniques for some linear algebraic functions are recalled [6-7]. A new justification is introduced dealing with geometrical transformations operations and their performance analysis under MorphoSys is proposed. These operations could be addition, subtraction or other operation supported in the MorphoSys processing elements (RC array). The speed of this mapping is calculated for different vectors and for different number of elements. The results compared with other processing systems. The M1 system yielded a speedup (with respect to number of cycles needed) of 8 and 10.5 for the 64-element translation and scaling algorithms respectively, as well as a speedup of around 37 for the 64-element rotation algorithm compared with the 80486. Finally, effort could be invested in trying to map other algorithms that make use of the mapped ones for more advanced algorithms for computer graphics.

**REFERENCES**

1. R. Maestre, F. Kurdahi, N. Bagherzadeh, H. Singh, R. Hermida, and N. Fernandez.Kernel Scheduling in Reconfigurable Computing. *Design and Test in Europe* (DATE'99), Munich, Germany, 1999.
2. N. Bagherzadeh, F. Kurdahi, H. Singh, G. Lu, M. Lee, and E. Filho. MorphoSys: An Integrated Reconfigurable System for Data-Parallel Computation-Intensive Applications. *IEEE Transactions on Computers*.
3. N. Bagherzadeh, F. Kurdahi, H. Singh, G. Lu, M. Lee, T. Lang, R. Heaton, and E. Filho. MorphoSys: An Integrated Reconfigurable Architecture. *NATO RTO Symposium of System Concepts and Integration*, Monterey, CA 1998.
4. N. Bagherzadeh, F. Kurdahi, H. Singh, G. Lu, M. Lee, and E. Filho, "MorphoSys: A Parallel Reconfigurable System", *Euro-Par 99*, Toulouse, France 1999.
5. N. Bagherzadeh, F. Kurdahi, H. Singh, G. Lu, M. Lee, and E. Filho. MorphoSys: A Reconfigurable Architecture for Multimedia Applications. *XI Brazilian Symposium on Integrated Circuit Design*, Rio De Janeiro, Brazil 1998.
6. I. Damaj, H. Diab. Performance Analysis of Some Geometrical Transformation Algorithms Using Reconfigurable Computing. *International Symposium on Innovation in Information & Communication Technology (ISIICT 2001)*, Amman, Jordan 2001.
7. I. Damaj, H. Diab. Performance Analysis of Extended Vector-Scalar Operations Using Reconfigurable Computing. *ACS International Conference of Computer Systems and Applications*, Beirut, Lebanon 2001.
8. I. Damaj, H. Diab. 2D and 3D Computer Graphics Algorithms under MorphoSys. *The 12$^{th}$ International Conference on Field Programmable Logic and Applications*, Montpellier, France 2002.

**Biography for Hassan Diab**

**Hassan Diab** received his B.Sc. in Communications Engineering, M.Sc. in Systems Engineering, and Ph.D. in Computer Engineering. He is a Professor of Electrical and Computer Engineering at the American University of Beirut, Lebanon. He has over 90 publications in international journals and conferences. His research interests include performance evaluation of parallel processing systems, reconfigurable computing, and simulation of parallel applications. Professor Diab is a Senior Member of IEEE and a Fellow of IEAust.

**Biography for Issam Damaj**

**Issam Damaj** received his B.Eng. in Computer Engineering, M.Eng. in Computer and Communications Engineering, and is currently a Ph.D. student at South Bank University, London. He is a Teaching Assistant in the School of Computing, Information Systems and Mathematics at South Bank University, London. His research interests include reconfigurable computing, fuzzy logic, and wireless communications security. He is a Member of the IEEE.